# Econo- and socio- physics based remarks on

# the economical growth of the World

Agata Angelika Rzoska

The Karol Lipiński Academy of Music, pl. Jan Pawła II 2, 50-043 Wrocław, Poland

e-mail: agatka.malwinka@gmail.com,

tel. 0048 694 502 564

**Abstract:**

It has been shown that the long term evolution of the Gross Product of the World after World War II can be well portrayed by the simply exponential function with the crossover at the year 1973, which coincides with the Oil Crisis onset. For such basic parameter as the S&P 500 index the single exponential behavior extends down to at least the mid of the nineteen century. It is notable that the detailed short-term insight focused on the last quarter of century revealed the emergence of the power like dependence, despite notable irregularities. However, such dependences can be introduced only when taking into account the behavior at reference-baselines years, clearly related to the society – important events in the past. The possible relationship to the growth/death evolution of microorganisms is also discussed.

**Key Words:** global product growth, econophysics, sociophysics, microbiology.

**Introduction**

In the last decades physics has entered new, interdisciplinary areas. In such approaches physics is most often used as a general analytic pattern for modelling problems so far understood mainly at the heuristic level, if not only via general discussions and reasoning. As an example can serve psychophysics which quantitatively investigates the relationship between physical stimuli and the sensations and perceptions they affect [1]. Medical Physics applies physical concepts, theories and methods to medicine or healthcare [2]. Sociophysics uses tools and concepts from physics to describe some aspects of social and political behavior and answers in the affirmative [1]. The emerging modern food science is the merge of the 'classical' applied food science and the soft matter physics, what creates the necessary surrounding for 'designing' innovative foodstuffs [3-5]. Probably, the most successful is the econophysics, applying theories and methods originally developed by physicists to solve problems in economics. Most often, it focuses on uncertainty issues, stochastic processes or nonlinear dynamics in finances. The term "*econophysics*" was introduced by R. Mantegna and H. E. Stanley [6] in the mid of nineties and later in the basic monography where he formulated key paths for this new discipline of knowledge [7]. It is notable that Stanley is also the author of the fundamental monography on the physics of critical phenomena and phase transitions [8], which methodology appeared to be particularly significant for econophysics.

Econophysics is extensively used for building advanced 'economic" theories, often using sophisticated physical tools for modelling [9]. However, to the best of the author knowledge, there are no reports regarding the most fundamental dependences in the 'general economics': the long term evolution of the global wealth or related productivity metrics [8, 10, 11]. Nowadays, this issue seems to be particularly important due to the decisive shift towards the planetary civilization and economy. Surprisingly, even the most prominent institutions, such as the International Monetary Fund, presents only rough plots for the time evolution of the Global Product or related properties [12-15].

The lack of the clear and justified parameterization of the most important plot describing the development of the global economy makes any general forecast beyond nowadays difficult. The target of the given report is approaching the solution of this Enigma. The analysis presented below is based on commonly available 'empirical' data, sourced in databases of such prominent institutions as the International Monetary Fund or the Earth Policy Institute [13-15].

**Results and Discussion**

The main part of Fig. 1 presents the plot illustrating the increase of the gross world product between years 1950 and 2014 [12, 13]. For the description of this dependence, the author assumed that the economy in the early fifties has to be very strongly influenced by World War II, one of the greatest tragedies of the mankind. One may expect that the great destruction of productive forces and the break of previous social links could notably 'reset' the World economy. After the short post-War depression period, the World economy converted into the new pattern: it focused on effectiveness and society important issues, with the democratic political background. All these were supported by almost unlimited demands for goods in rebuilding Europe and the rapid growth in USA [16]. As the onset of this period one can consider the year 1948, 3 years after World II. In fact this was the terminal of first post-World reconstruction plans in Europe. Hence, taking year 1948 and the related world product as the reference (*ref.*) may be a reasonable assumption for the analysis. The inset in Fig. 1 shows that using such baseline for 'empirical' data from the main part of the plot and subsequently applying the semi-log scale the simple exponential dependence emerges:

$$\Delta V(\Delta Y) = y_0 \exp(\Delta Y \times V_{act.}) \qquad (1)$$

where: $\Delta Y = Y - Y_{ref.}$ is for the distance from the reference year $Y_{ref.} = 1948$ and for the gross world product $\Delta V = V - V_{ref.}$ and $V_{ref.} = V(1948) = 4$

It is worth recalling that for semi-log plots the exponential behavior manifests via the linear function, easily "eye visible" in the inset in Fig. 1, namely:

$$\Delta V(\Delta Y) = y_0 \exp(\Delta Y \times V_{free}) \rightarrow \ln \Delta V = \ln y_0 + V_{free} \times \Delta Y \rightarrow a + bx \qquad (2)$$

where: $a = \ln y_0$ is the intercept and $b = V_{free}$ is for the slope of the line ($x = \Delta Y$, $y = \ln \Delta V$). Coefficients $a$ and $b$ ($y_0 = e^a$) can obtained be from the linear regression fit. It is notable that one linear dependence in the inset (eq. (2)) describes empirical results up to the year 1973 and then the evolution cross over to the another one linear description in the inset. The latter continues until nowadays. It is notable that in October 1973 the cartel of key oil producers OPEC (*Organization of the Petroleum Exporting Countries*) proclaimed embargo for the delivery of the petroleum oil. In 1974 the price of oil rose from $3 per barrel to $12 - $15, i.e. 400% - 500% occurred! The Oil Crisis has continued for decades [17, 18]. The behavior revealed in the inset in Fig. 1 may indicate that this event created permanent "new constraints", which has governed the World Economy until now (!). Notable is the lack of next "Oil Crises" hallmarks in the plot, even such notably as the one in 1979 associated with the crisis in Iran [19]. This can mean that the pattern of economic constraints created in 1973 are still valid.

Simple exponential relations are relatively common in physics. They are mainly associated with the behavior of non-interacting or weakly interacting " 'species". In the opinion of the author particularly worth recalling is the Barus equation describing the pressure (*P*) dependence of such properties as diffusion ($D(P)$), relaxation time ($\tau(P)$) or viscosity ($\eta(P)$), for instance [20, 21]:

$$\tau(P) = \tau_o \exp(P \times V_{free}) \qquad (3)$$

where: $V_{free}$ is the free volume available for an orientation, translation … of basic species/units in the given system. The Barus eq. (3) can be considered as the pressure counterpart of the famous Arrhenius dependence $\tau(T) = \tau_o \exp(E_a/RT)$: $E_a$ is the process activation energy and $R$ the gas constant [21, 22].

The formal similarity of the Barus equation (*see below*) [20, 21] to the dependence describing empirical data in the inset in Fig. 1 (see also eq. 1) makes it possible to speculate that the impact Oil Crisis in 1973 decreases notably the '*free volume*' for economic activities: $V_{free}$:

$[V_{free}(1973-2014)/V_{free}(1950-1973)] \times 100\% \approx 270\%$ (see the caption of Fig. 1).

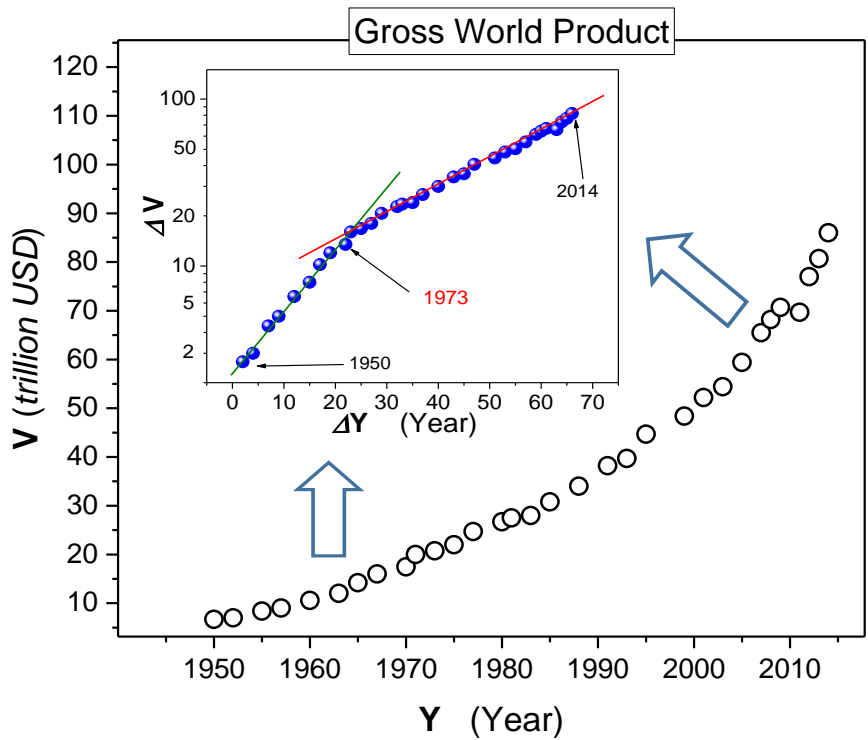

Fig. 1 The increase of value (*V*) of the Gross World Product in the - World War II period, until nowadays. Data are taken from refs. [12-15]. The main part present data in the commonly applied scale. The inset shows the same 'experimental' data taking the year 1948 with the product ca. 4 trillion USD as references (ref.), namely: $\Delta Y = Y - Y_{ref.}$

and $\Delta V = V - V_{ref.}$. The applied semi-log scale reveals the basic exponential relation: $\Delta V(\Delta Y) = y_0 \exp(\Delta Y \times V_{free})$ via the simple linear dependence at the plot. The subsequent linear regression analysis yielded: $y_0 = 2.25$ and $V_{free} = 0.09 \pm 0.02$ (for years 1950 – 1973) and $y_0 = 1.9$ and $V_{free} = 0.035 \pm 0.0004$ (for years 1973– 2014).

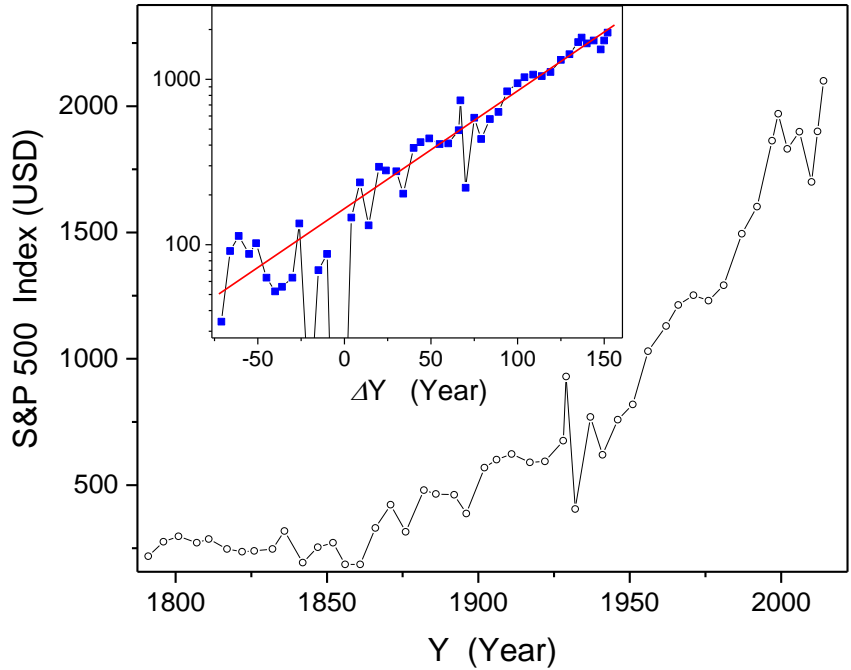

Fig. 2  Changes of the S&P 500 index between years 1791 and 2014. Data are taken from ref. [12-15]. The main part presents 'experimental' data in the commonly applied way. The inset present the same data in the semi-log scale with, assuming scale with reference baselines, namely: $\Delta Y = Y - Y_{ref.}$ and D(S&P) = S&P - S&P$_{ref}$. The resulted linear dependence visualize the appearance of a simple exponential behavior portrayed by: $S\&P(\Delta Y) = sp_0 \exp(\Delta Y \times V_{free.}^{SP})$, with $sp_0 = 158 \pm 20$ and $V_{free.}^{SP} = 0.017 \pm 0.02$. Data were fitted above the reference year $Y > Y_{ref.} = 1862$. Below this 'threshold time"

the red curve in the inset shows the extrapolation. For fitting the linear regression was used.

Fig. 2 presents the similar way of analysis as above applied for the S&P 500 index, extrapolating down to the year 1791, i.e. the onset of economy in the new independent country: USA. The S&P 500 (*Standard&Poor's 500*) is an American stock market index based on the market capitalizations of 500 largest companies listed on the NYSE (*New York Stock Exchange*) or NASDAQ (*National Association of Securities Dealers Automated Quotations*) [11, 23]. For 'empirical' data presented in the main part of Fig. 2 S&P 500 indexes were normalized to the year 2000 [12, 13]. For times before the appearance of NYSE and NASDAQ, key companies were selected and possible values of indexes were estimated [12, 13]. The quantitative analysis presented in the inset in Fig. 2 reveals a clear linear dependence when using the semi-log plot, but only if following reference baselines are taken into account: $\Delta Y = Y - Y_{ref.}$ and $\Delta(S\&P) = S\&P - S\&P_{ref.}$, where $Y_{ref.} = 1862$ and $S\&P_{ref.} = 184$. The last number is related to the mean index in the decade prior to the year 1862. As shown in the inset in Fig. 2 the evolution of the index can be portrayed via:

$$S\&P(\Delta Y) = sp_0 \exp\left(\Delta Y \times V_{free.}^{SP}\right) \tag{4}$$

with $sp_0 = 158 \pm 20$ and $V_{free.}^{SP} = 0.017 \pm 0.02$ were determined from the linear regression fit.

In the opinion of the author year 1862 may be not accidental, since it coincides with the beginning of the American Civil War. Although during this War the economy of Confederate States collapsed, the economy of Union (North States) flowered and its value at least doubled at the end of the War. The end of the Civil War created conditions to the unrestricted boost of USA economy. At the onset of 20$^{th}$ century USA reached the level of the World greatest 'economy' [24].

It is interesting that the same pattern of S&P500 index evolution seems to be valid from the beginning 19$^{th}$ to nowadays. This may suggest that restrictions/constraints for the World - largest corporations in year 2014 and in 19$^{th}$ century (the time of the almost unrestricted capitalism !) are similar.

Fig. 3 discusses once more the world global product, but focusing on the last 25 years (basing on detailed data from refs. [12, 13]). The main part of the plot shows the basic and commonly used dependence. The strongly irregular behavior is visible. Its parameterization seems to be difficult, if possible at all. The red line for years 2014+ shows the official extrapolation of the IMF up to year 2020. It is expected (by IMF) that the global economy can approach the enormous value ~ 100 trillion USD. However, no explanation for this prediction was given [12-15]. The application of the semi-log plot, as in Figs. 1 and 2, did not yield any conclusive behavior for data from main part of Fig. 3. However, an interesting picture emerges when using the log-log plot, with the reference-baseline related to the year 1989:

$$\Delta(GPB) = p_{ref} \Delta Y^{\phi} \text{ and then: } \log_{10}[\Delta(GBP)] = \log_{10} p_{ref} + \phi \log_{10} \Delta Y = a + bx \qquad (5)$$

where: $\Delta(GPB) = GPB - GPB_{ref.} (for\_Y = 1989)$, $\Delta Y = Y - 1989$. For the linear dependence in the inset (red line): $x = \log_{10} \Delta Y$ and $y = \log_{10} \Delta(GBP)$, the intercept $a = \log_{10} p_{ref.}$ (and then $p_{ref.} = 10^a$) and the slope of the line $b = \phi$. Values of relevant parameters in eq. (5): the exponent $\phi = 1.2 \pm 0.1$ and the prefactor $p_{ref.} = 1.75 \pm 0.1$ where obtained using the linear regression method.

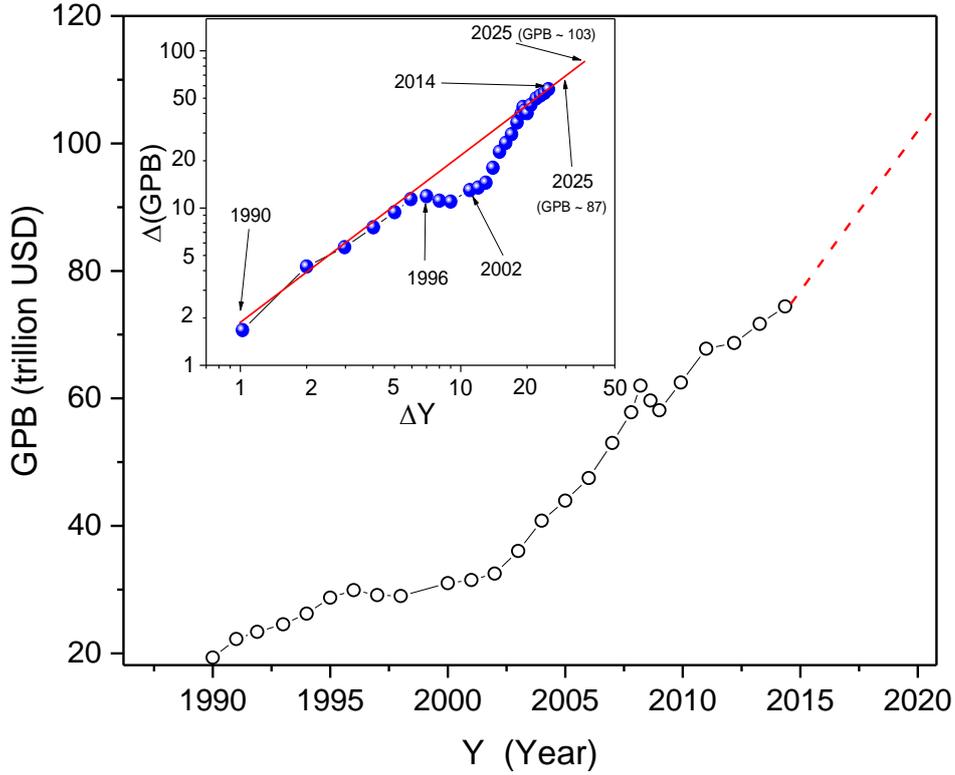

Fig. 3 The global product (GPB) in the last quarter of century, based on data from ref. [6, 7]. The main plot shows the common way of presentation and the inset the log – log plot assuming, reference/onset values: $\Delta Y = Y - Y_{ref.}(=1989)$ and $\Delta(GPB) = GPB - GPB_{ref.}(for\_Y = 1989)$. The emerging linear behavior indicates the parameterization via: $\Delta(GPB) = p_{ref}\Delta Y^\phi$, with $p_{ref.} = 1.75 \pm 0.1$ and $\phi = 1.2 \pm 0.1$.

It is notable that the parameterization presented in the inset in Fig. 3 enables 'justified' extrapolations, yielding for the year 2020 the value ca. 13 % lesser than expected so far. In the opinion of the author, the reference to the year 1989 may be not accidental. In this year first in Poland and next in other "people democracy" countries communism collapsed. These countries, including Russia, entered the World of *true economy* [16]. Huge demands for commodities in 'former communists' countries strongly stimulated the economy of western countries (EC, USA, …). Hence, 1989 may be a kind of the 'onset year' for the World economy.

The inset in Fig. 3 reveals the power-type dependence $\Delta(GPB) = p_{ref}\Delta Y^{\phi}$ (eq. 5) with a notable distortion in the period 1996–2008. It started in the year 1996, the same in which a deep economical crises in Russia and few other east-European countries started. It was caused mainly by non-clear political structures and deeply non-proper *privatization/commertialization* of previously state-own companies [16]. Years 1996–2001 are also related to the so called *dot-com bubble* [25] associated with the boost of virtual companies linked solely to the Internet and their subsequent collapse. Subsequently, starting slowly from the year 2003 the world economy moved towards the 'blooming" pattern from years 1990 - 1996. This was the time of 'new Putin order" in Russia as well as entering key European countries (Poland, Czech, Slovak, Hungary, …..) into the European Community. China reached the unusually large growth rates each year. It is notable that in the inset the "21$^{st}$ great bank related crisis" near 2010 [16] is only a small distortion with the negligible impact on the general trend in the inset in Fig. 3.

Concluding, basing on the way of analysis used for experimental data in physics clear trends describing the World ('planetary') economy have been found. Such dependences make a reliable forecasts of future trends possible. Resulted values, given in the inset in Fig. 3, a smaller than official estimations of IMF.

The question arises if the exponential increase of the world economy can be continued in subsequent decades of the 21$^{st}$ century. Some degree of independence of "economic beings", their appearance/disappearance and their simple exponential portrayal (discussed above) may suggest the similarity of the economic growth to the life cycle of bacteria/microorganisms in the system of defined and restricted resources. It is worth stressing that the economy has reached the planetary level nowadays. Moreover, for many natural resources the terminal point is visible. Parallel, new and extremely difficult political and social problems have emerged. This is assisted by the global population increase, although occurring beyond the most developed areas. All these creates new and 'heavy' constraints/problems for the most economically productive part of the World. In the opinion of the author these factors can negatively influence the global growth rates.

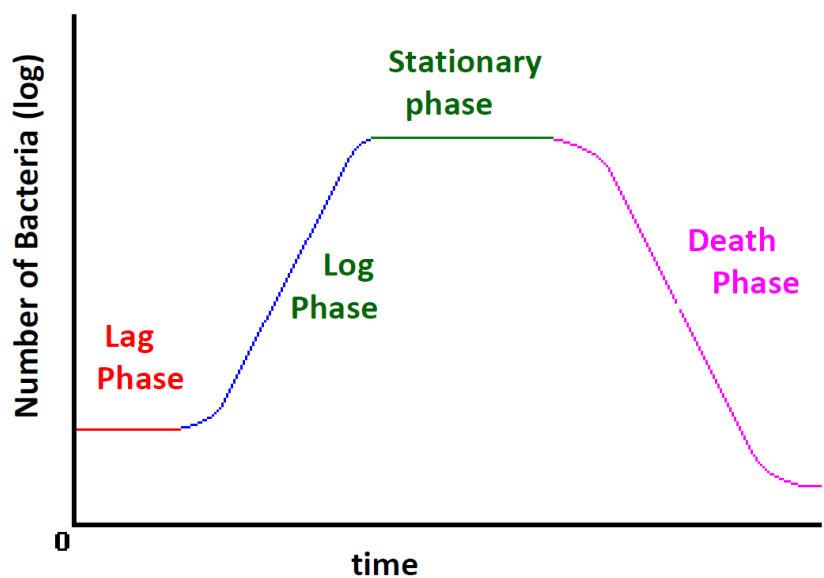

Fig. 4  The growth&death cycle for bacteria/microorganisms (prepared basing on ref. [26]). After the stable population period (lag phase) the strong increase described by the single exponential dependence occurs (the log-phase"). Next, the population stabilizes (stationary phase). Finally, the number of microorganisms decreases also following the exponential pattern (but with the negative sign). *Exponential patterns for the growth and death parts manifest mainly via linear functions due to the applied semi-log scale.*

Consequently, the question arises if the World economy approaches the "stationary phase" known for the life-cycle of microorganisms population (see Fig. 4). For such tendency may indicate a slower than exponential one evolution for the last decades as noted in Fig. 3. In the 'stationary phase", without economic growth, new and serious social, political and economical problems can appear. All these can terminate in the decay/death phase for economy (and civilization ?). It also notable that when considering the general population growth of some countries, for instance Poland or Japan, it also start to resemble the pattern from Fig. 4. The question arises if such dependence extends for the global level in the future.

In the opinion of the author all these may on seemingly inevitable phase of stabilization and the further decline, first for the most developed part of the World and next globally. The emigration to the most developed countries can stop this hypothetical process locally and temporarily.
It is worth recalling here that for a colony of microorganism the final stage is the complete disappearance.

Hence, the question whether if the emerging 'general civilization problem' of the possible inevitable decay of civilization can be solved. For a colony of microorganisms in a bio-system two scenarios can be proposed:

**(i)** delivering of extra food, water, oxygen, light,… within the given restricted areas ('*more foods in the same restricted area*")

**(ii)** removing space restrictions and opening new areas for "conquering" and developing ('*new lands available*').

In the case of the mankind point this can be implemented as follows:

- For point (**i**): *new breakthrough findings/discoveries well beyond the current state-of-the-art*: *for instance the solution of energy and foods limitations problems matched with political and social 'innovative' approaches.*

- For point (**ii**): *the 'phase transition" toward the 'solar civilization", where 'infinite" resources can be available and the 'infinite space" of the Solar system becomes open for the exploration.*

In summary, the report proposes the new discussion of the past and nowadays time of the global economy. It recalls econonophysics and sociophysics as disciplines within which the effective parameterization of trends is possible. Finally, possible future trends are discussed.

**Acknowledgements**

The author is grateful to Aleksandra Drozd-Rzoska and Sylwester J Rzoska for the support, discussion and redaction consultations. Consequently, this research was supported by the grant ref. UMO-2011/03/B/ST3/02352 (NCN, Poland).